\documentstyle[twocolumn,11pt]{article}

\def\spose#1{\hbox to 0pt{#1\hss}}
\def\simlt{\mathrel{\spose{\lower 3pt\hbox{$\mathchar"218$}}
     \raise 2.0pt\hbox{$\mathchar"13C$}}}
\def\simgt{\mathrel{\spose{\lower 3pt\hbox{$\mathchar"218$}}
     \raise 2.0pt\hbox{$\mathchar"13E$}}}
\date{}
\begin{document}

\title{PROBING THE ``DARK AGE'' WITH NGST}   

\author{Martin J. Rees \\ Institute of Astronomy,
Madingley Road, Cambridge, CB3 OHA}

\maketitle

\begin{abstract}
  At redshifts between 5 and 20, stars and `subgalaxies' created the
first heavy elements; these same systems (together perhaps
with  `miniquasars') generated the UV radiation that ionized the IGM,
and maybe also the first magnetic fields.  The history of the universe
during this crucial formative stage is likely to remain highly
uncertain until the launch of NGST.
\end{abstract}

\section{Introduction}
    The Universe literally entered a dark age about half a million
years after the big bang: the primordial radiation then cooled below
3000K and shifted into the infrared.  Darkness persisted until the
first non-linearities developed, and evolved into stars, galaxies or
black holes that lit the Universe up again.  Darkness will again
gradually descend after $10^{14}$ years, when even the slowest-burning
stars have ended their lives (if there has not been a big crunch in
the meantime). I won't be looking that far ahead, But it is important,
in discussing NGST, to attempt to forsee what progress will have been
made by 2007, and what will still be the key ${\rm enigmas}$.

   We will by then surely have elucidated the history of star
formation, galaxies and clustering back to $z = 5$. The evolution of
galaxies of all morphological types will have been clarified by
further Hubbble Deep Fields, together with follow-up spectroscopy from
the new generation of 10 metre telescopes. Also, full analysis of
absorption features in quasar spectra (the Lyman forest, etc), will
probe the clumping, temperature, and composition of diffuse gas on
galactic (and smaller) scales, at least back to $z=5$.

    By 2007 detailed sky maps of the microwave background (CMB)
temperature (and perhaps its polarization as well) will offer direct
diagnostics of the initial linear fluctuations from which the
present-day large-scale structure developed. However, these
measurements will not directly probe the small angles that are
relevant to the (subgalactic-scale) structures, which, in any
hierarchical (`bottom up') scenario would be the first non-linearities
to develop.

    We may by 2007 know the nature of the dark matter, and how it
clusters gravitationally; computer simulations will incorporate gas
dynamics and radiation in a sophisticated way.
 
      But these advances will still leave us very uncertain about the
whole era from $10^7$ to $10^9$ years, and about the faint precursors of
galaxies -- the first stars, the first supernovae, the first heavy
elements; and how and when the intergalactic medium was reionized.  We
will probably still be unable to compute crucial things like the star
formation efficiency, feedback from supernovae. etc -- processes that
current models for galactic evolution are forced to parametrise in a
rather ad hoc way.

\section{Cosmogonic\\ Preliminaries}
  Most detailed studies of structure formation have focused on the
cold dark matter (CDM) model. Even if its details prove incorrect, it
is still a useful `template' whose main features apply generically to
any `bottom up' model for structure formation.  There is no minimum
scale for gravitational aggregation of the CDM. However, the baryonic
gas does not `feel' the very smallest clumps: pressure opposes
condensation on scales below the baryonic Jeans mass.

    Even though the low-mass structures that form first have a higher
density than galactic halos that form later, their associate
gravitational potential wells are shallower. In contrast to galaxies
(virial velocities $\sim 300$ km/sec), and clusters (up to 1000 km/sec) the
first structures in which baryons would condense have a scale set by
the Jeans mass at the minimum temperature where cooling can occur.
Atomic hydrogen cooling is efficient at temperatures above $10^4$K
(virial velocities $\simgt 10$ km/sec); this yields a characteristic
mass of $10^9 ((1 + z)/10)^{-3/2} M_\odot$ . Cooling by molecular hydrogen may
allow even smaller systems to form.  Feedback effects are very
important in these shallow potential wells -- for instance, one
supernova could eject many thousand times its own mass, and even
photoionization by the first O and B stars could dislodge gas from
these shallow potential wells.

   The IGM remained predominantly neutral until a sufficient number of
`subgalaxies' above this characteristic mass had gone non-linear to
provide the requisite O-B stars (or accreting black holes) that
photoionized the IGM.
    How many of these `subgalaxies' formed, and how bright each one
would be, depends on another big uncertainty: the IMF for 
Population III stars.  They form in an unmagnetised medium of pure H
and He bathed in background radiation which may be hotter than 50 K
when the action starts (at redshift $z$ the temperature is of course
2.7(1 + z) K).  Would these conditions favour a flatter or a steeper
IMF than we observed today?  This is completely unclear: the density
may become so high that fragmentation proceeds to very low masses
(despite the higher temperature and absence of coolants other than
molecular hydrogen); on the other hand, massive stars may be more
favoured than at the present epoch. Indeed, fragmentation could even
be so completely inhibited that the first things to form are
supermassive holes.
   
 The gravitational aspects of clustering can be computed with
convincingly high resolution. So also, now, can the dynamics of the
baryonic (gaseous) component -- including shocks and radiative
cooling. But the huge dynamic range of the star-formation process
cannot be followed deductively, so the nature of the simulation
changes as soon as the first stars (or other compact objects) form.
The first stars exert crucial feedback -- the remaining gas is heated
by ionizing radiation, and perhaps also by an injection of kinetic
energy via winds and even supernova explosions -- and this depends on
the IMF, and on further uncertain physics.

  So three major uncertainties  are:

(i) What is the IMF of the first stellar population? The high-mass
stars are the ones that provide efficient (and relatively prompt)
feedback. It plainly makes a big difference whether these are the
dominant type of stars, or whether the initial IMF rises steeply
towards low masses (or is bimodal), so that very many faint stars form
before there is a significant feedback.

(ii) Quite apart from the uncertainty in the IMF, it is also unclear
what fraction of the baryons that fall into a clump would actually be
incorporated into stars before being re-ejected. The retained fraction
would almost certainly depend on the virial velocity: gas more readily
escapes from shallow potential wells.

(iii) The influence of the early stars depends on whether their energy
is deposited locally or penetrates into the medium that is not yet in
contracting systems. The UV radiation could, for instance, be mainly
absorbed in the gas immediately surrounding the first stars, so that
it exerts no feedback on the condensation of further clumps -- the
total number of massive stars or accreting holes needed to build up
the UV background, and the possible concomitant contamination by heavy
elements, would then be greater.

       Perhaps I'm being pessimistic, but I doubt that either
observations or theoretical progress will have eliminated these
uncertainties about the `dark age' by the time NGST flies.  Later
speakers at the conference, especially Drs Loeb, Ferrara and Madau,
will be addressing some aspects of these issues. I will merely try to
highlight some key questions that NGST could address.

\section{The Epoch of Ionization Breakthrough}
\subsection{Why is the breakthrough epoch important?}
     
Quasar spectra tell us that the IGM is almost fully ionized back to
$z=5$, but we do not know when the universe in effect became an HII
region.  This question must be answered before we can properly
interpret angular fluctuations in the CMB radiation.  If the
intergalactic medium is essentially fully ionized back to a redshift
$z_i$, then the scattering opacity scales roughly as $(1+ z_i) ^{3/2}$. Even
when this optical depth is far below unity, the ionized gas
constitutes a `fog' that attenuates the fluctuations imprinted at the
recombination era; the fraction of photons that are scattered at $<z_i$
then manifest a different pattern of fluctuations, characteristically
on larger angular scales. This optical depth is consequently one of
the parameters that can in principle be determined from CMB anisotropy
measurements. It is feasible to detect a value as small as 0.1 --
polarization measurements of the kind expected from Planck-Surveyor
may allow even greater precision, since the scattered component would
imprint polarization on angular scales of a few degrees, which would
be absent from the Sachs Wolfe fluctuations on that angular scale
originating at $t_{\rm rec}$.

   The thermal history of the IGM beyond $z= 5$ is relevant to the
modelling and interpretation of the absorption spectra of quasars at
lower redshifts. The recombination and cooling timescales for diffuse
gas are comparable to the cosmological expansion timescale. Therefore
the `texture' and temperature of the filamentary structure responsible
for the lines in the `forest' carry fossil evidence of the thermal
history at higher redshifts.

\subsection{Why is \mbox{\boldmath $z_i$}  so uncertain?}
   Even if we knew exactly what the initial fluctuations were, and
when the first bound systems on each scale formed, the efficiency of
UV generation in the IGM depends on OB star formation, possibly on
early black hole formation, and also on whether UV photons can escape
from the dense gas around their points of origin. These processes are
so poorly understood that the breakthrough redshift $z_i$ is uncertain by
a factor of 2, even if we postulate a `standard' IMF; once we admit a
possibly different IMF , the uncertainty widens still more.  This can
be easily seen as follows:

      Ionization breakthrough requires 1-10 photons for each baryon in
the IGM.  An OB star produces $10^4$ -- $10^5$ ionizing photons for each
constituent baryon, so the requisite UV could be supplied by $10^{-3}$ of
the baryon turning into stars with a standard IMF. We can then
contrast two cases:

(A).  If the star formation were efficient, in the sense that all the
baryons that `went non-linear', and fell into a CDM clump larger than
the Jeans mass, turned into stars, then only the rare 3-$\sigma$ peaks in the
initial fluctuation spectrum would suffice.  On the other hand:

(B). Star formation could plausibly be so inefficient that less than
1 percent of the baryons in these pregalactic systems condense into
stars, the others being expelled by stellar winds, supernovae, etc.,
In this case, production of the necessary UV would have to await the
collapse of more typical peaks (1.5-$\sigma$, for instance).

      A 1.5-$\sigma$ peak has an initial amplitude only half that of a 
3-$\sigma$ peak, and would therefore collapse at a value of $(1 + z)$ that was
lower by a factor of 2.  For plausible values of the fluctuation
amplitude this would change $z_i$ from 15 (scenario A) to 7 (scenario
B). There are of course other complications, stemming from the
possibility that many of the UV photons may be reabsorbed locally;
moreover in Scenario B the formation of sufficient OB stars might have
to await the build-up of larger systems in which stars would form more
efficiently.

      If the IMF were biased towards low-mass stars, the situation
resembles inefficient star formation with respect to ionization.
However, there is the possibility that, before enough UV as been
generated, a substantial fraction of the baryons could be condensed
into low mass stars. This population could be sufficient to contribute
to the MACHO lensing events (see section 5).

      Note that scenarios A and B would have interestingly different
implications for the formation and dispersal of the first heavy
elements. If B were correct, there would be a large number of
locations whence heavy elements could spread into the surrounding
medium; on the other hand, scenario A would lead to a smaller number
of more widely spaced sources.

\subsection{AGNs at high \mbox{\boldmath $z$}?} 
      By the epoch $z = 5$, some structures (albeit perhaps only
exceptional ones) must have attained galactic scales. Massive black
holes (manifested as quasars) accumulate in the deep potential wells
of these larger systems. Quasars may dominate the UV background at 
$z <3$: if their spectra follow a power-law, rather than the typical
thermal spectrum of OB stars, then quasars are crucial for the second
ionization of He. AGN formation may require virialised systems with
large masses and deep potential wells (cf Haehnelt and Rees 1993); if
so, we would naturally expect the UV background at the highest
redshifts to be contributed mainly by stars in `subgalaxies'. However,
this is merely an expectation; it could be, contrariwise, that black
holes readily form even in the first $10^8 M_\odot$ CDM condensations (this
would be an extreme version of a `flattened' IMF), Were this the case,
the early UV production could be dominated by black holes. This would
imply that the most promising high-$z$ sources to seek with NGST would
be miniquasars, rather than `subgalaxies'. It would also, of course,
weaken the connection between the ionizing background and the origin
of the first heavy elements.

 \subsection{Detectability of `pregalactic' UV sources and
   determination of \mbox{\boldmath $z_i$}} 
The detectability of these early-forming
   systems, of subgalactic mass, depends 
on whether (A) or (B) (section 3.2) is
 nearer the truth.  This is discussed in a recent paper by Jordi
   Miralda Escud\'e and myself (1998). There are already some
   constraints from the Hubble Deep Field, particularly on the number
   of `miniquasars' (Haiman, Loeb and Madau 1998).  Objects down to AB
   mag of 31 could be detected from the ground by looking at a field
   behind a cluster where there might be gravitational-lens
   magnification; but firm evidence is likely to await NGST.

\subsection{Distinguishing between objects with \mbox{\boldmath $z> z_i$} and 
\mbox{\boldmath $z< z_i$}}
     The blanketing effect due to the Lyman alpha forest -- known to
be becoming denser towards higher redshifts, and likely therefore to
be even thicker beyond $z=5$ -- would be severe, and would block out the
blue wing of Lyman alpha emission from a high-$z$ source.  Such objects
may still be best detected via their Lyman alpha emission even though
the absorption cuts the equivalent width by half. But at redshifts
larger than $z_i$ -- in other words, before ionization breakthrough --
the Gunn-Peterson optical depth is so large that any Lyman alpha
emission line is blanketed completely, because the damping wing due to
IGM absorption spills over into the red wing (Miralda Escud\'e 1998).
This means that any objects detectable beyond $z$ would be found by a
discontinuity at the redshifted Lyman alpha frequency. The Lyman alpha
emission line itself would not be detectable (even though this may be the
easiest feature to detect in objects with $z<z_i$).

\section{Detecting Very Distant Supernovae with NGST}
        It is straightforward to calculate how many supernovae would
have gone off, in each comoving volume, as a direct consequence of
this output of UV and heavy elements if the reheating and ionization
were due to OB stars (cf Haiman and Loeb 1997, Gnedin and Ostriker
1997, Songalia 1997, Cowie 1998): there would be one, or maybe
several, per year in each square arc minute of sky (Miralda-Escud\'e and
Rees 1997).  The uncertainty depends partly on the redshift and the
cosmological model, but also on the uncertainties about the UV
background, and about the actual production of heavy elements. (These
may be overestimated because the heavy element distribution is
`patchy', and concentrated in the overdense regions that yield
high-column density absorption features. On the other hand, they may
be underestimated if most are concentrated in the sources; moreover,
if most of the UV is absorbed near its source, then more production is
needed to generate the required intergalactic ionization.)

   These high-$z$ supernovae would be primarily of Type 2.  The typical
observed light curve has a flat maximum lasting 80 days. One would
therefore (taking the time dilation into account) expect each
supernova to be near its maximum for nearly a year. It is possible
that the explosions proceed differently when the stellar envelope is
essentially metal-free, yielding different light curves, so any
estimates of detectability are tentative. However, taking a standard
Type 2 light curve (which may of course be pessimistic), one
calculates that these objects should be approx 27th magnitude in J and
K bands even out beyond $z = 5$.  The detection of such objects would be
an easy task with the NGST. With existing facilities it is
marginal. The best hope would be that observations of clusters of
galaxies might serendipitously detect a magnified
gravitationally-lensed image from far behind a cluster.
   
     The first supernovae could be important for another reason: they
may generate the first cosmic magnetic fields. Mass loss (via winds or
supernovae permeated by magnetic flux) would disperse magnetic flux
along with the heavy elements. This flux, stretched and sheared by
bulk motions, can be the `seed' for the later amplification processes
that generate the larger-scale fields pervading disc galaxies.

   (Incidentally, it is now clear that the afterglows of gamma-ray
bursts are 100 times brighter than supernovae. The rate, however, is
low -- even though it could exceed that of the bursts themselves if
the gamma rays are more narrowly beamed than the slower-moving ejecta
that cause the afterglow.  Detection of an afterglow beyond $z=5$ would
offer a marvellous opportunity to obtain a high-resolution spectrum of
intervening absorption features.)

 \section{Where are the Oldest Stars?}  
The efficiency of early mixing is important for the interpretation of
stars in our own galaxy that have ultra-low metallicity -- lower than
the mean metallicity that would have been generated in association
with the UV background at $z > 5$ (cf Norris 1994).  If the heavy
elements were efficiently mixed, then these stars would themselves
need to have formed before galaxies were assembled. To a first
approximation they would cluster non-dissipatively; they would
therefore be distributed in halos (including the halo of our own
Galaxy) like the dark matter itself. More careful estimates slightly
weaken this inference, This is because the subgalaxies would tend,
during the subsequent mergers, to sink via dynamical friction towards
the centres of the merged systems. There would nevertheless be a
tendency for the most extreme metal-poor stars to have a more extended
distribution in our Galactic Halo, and to have a bigger spread of
motions. This is another project where NGST could be crucial,
especially if it allowed detection of halo stars in other nearby
galaxies.

   The number of such stars depends on the IMF. If this were flatter, there
would be fewer low-mass stars formed concurrently with those that produced the
UV background. If, on the other hand, the IMF were initially steeper, there
could in principle be a lot of very low mass (macho) objects produced at high
redshift, many of which would end up in the halos of galaxies like our own.

\section{Summary}
  
      Perhaps only 5 per cent of star formation occurred before
$z=5$. But at the conference on NGST held last year at Goddard Space
Flight Center, Alan Dressler, in his concluding lecture, rightly
emphasised that these early stars were important, just as were the
first 5 percent of humans. There is still a variety of models for
cosmic structure, that seem consistent with the properties of our
universe at the current epoch.  Large-scale structure may be
elucidated within the next decade, by ambitious surveys (2-degree
field and Sloan) and studies of CMB anisotropies. But there will still
be uncertainty about how the present structures emerged, and
especially about the efficiency and modes of star formation in early
structures on subgalactic scale.
 
    NGST will have many roles. But in deciding the tradeoffs between
aperture, waveband coverage, and image quality, it is important to
optimise this unique chance to elucidate the formative stages of
cosmic structure.

\section*{Acknowledgements}
I am grateful to my collaborators, especially Zoltan Haiman, Avi Loeb,
Jordi Miralda-Escud\'e and Max Tegmark, for discussion of the topics
described here.

\section*{References}

\def\ref1{\par\noindent\hangafter=1\hangindent=20pt}
\ref1 Cowie, L. 1998 preprint
\ref1 Gnedin, N. \& Ostriker, J.P. 1997 ApJ 486, 581
\ref1 Haehnelt, M. and Rees, M.J. 1993\hfill\break MNRAS 263, 168
\ref1 Haiman, Z. and Loeb, A. 1997 ApJ 483, 21
\ref1 Haiman, Z., Loeb, A. and Madau, P. 1998 ApJ  (submitted)
\ref1 Miralda-Escud\'e, J. 1998 ApJ (in press)
\ref1 Miralda-Escud\'e, J. and Rees, M.J. 1997 ApJ 478, L57
\ref1 Miralda-Escud\'e, J. and Rees, M.J. 1998 ApJ (in press)
\ref1 Norris J.E. 1994 ApJ 431, 645
\ref1 Songalia, A. 1997 ApJ 490, L1
\end{document}